\documentclass[%
 preprint,
 floatfix,
 amsmath,amssymb,
 aps,
]{revtex4-2}
\pdfoutput=1
\usepackage{graphicx}% Include figure files
\usepackage{dcolumn}% Align table columns on decimal point
\usepackage[normalem]{xcolor}
\usepackage{soul}
\usepackage{bm}% bold math
\usepackage{amssymb}
\usepackage{amsthm}
\usepackage{amsmath}
\usepackage{miller}%For Miller Indices
\usepackage{textcomp}
\usepackage{gensymb}
\usepackage{siunitx}
\usepackage[utf8x]{inputenc}
\usepackage{multirow}
\usepackage{natbib}
\newcommand*{\fig}[1]{Figure~\ref{fig:#1}}
\newcommand*{\sect}[1]{Section~\ref{sec:#1}}

\newcommand*{\tab}[1]{Table~\ref{tab:#1}}

\begin{document}

%\preprint{APS/123-QED}

\title{Dislocations as natural quantum wires in Diamond}% Force line breaks with \\

\author{Sevim Polat Genlik}
 \altaffiliation{polatgenlik.1@osu.edu}%Lines break automatically or can be forced with \\
\author{Roberto C. Myers}
\author{Maryam Ghazisaeidi}
 \email{ghazisaeidi.1@osu.edu}
\affiliation{%
Department of Materials Science at The Ohio State University, Columbus, Ohio 43210, USA.
}%https://www.overleaf.com/project/6242720cbbbe3521243f7e36

\date{\today}% It is always \today, today,
             %  but any date may be explicitly specified
\begin{abstract}

We study the electronic properties of the glide set of dislocations in diamond from first principles using  hybrid exchange correlation functionals and find that the atomic-scale dislocation core states give rise to a prototypical one-dimensional (1D) band structure, i.e. natural quantum wires. The position and character of the core states varies strongly with local structure, where mixed dislocations with dangling bonds exhibit a 1D metallic band with a characteristic 1D density of states ($1/\sqrt{E})$. This 1D Fermi gas is spatially localized to single atomic diameter orbital chain along the dislocation core.  When the dangling bonds within the core are reconstructed, the 1D metallic band disappears. In contrast, pure edge dislocations in diamond reveal a 1D semiconductor with a direct band gap of 3.0 eV. These calculations provide a possible explanation to the long standing observation of a blue luminescence band correlated with dislocations in diamond. This opens the door to using dislocations as 1D quantum phases with functional (electronic and optical) properties arising from the atomic-scale core states.
\end{abstract}
\maketitle

%\tableofcontents

\section{\label{sec:level1}INTRODUCTION}

Low dimensional quantum materials have attracted significant attention because of their distinct properties from bulk. Despite extensive research on low dimensional materials,  major challenges in using them remain related to their fabrication and storage\cite{peaker2013low}. 
Dislocations--1D defects in crystals-- are similar to low dimensional materials, in the sense that the different bonding environment at their core leads to local and distinct properties than those in their surrounding crystal. They can be used as  templates for creating conducting nanowires in insulating materials\cite{nakamura2003conducting, ikuhara2009nanowire,tokumoto2009fabrication,amma2010electrical} or ferromagnetic nanowires in antiferromagnetic materials\cite{sugiyama2013ferromagnetic}. Moreover, dislocations are embedded in the solid, and as such are environmentally protected by their host. This is in contrast to existing low dimensional quantum materials-- such as metallic or semiconducting nanowires-- which degrade in short time due to their instability\cite{zhou2014long}.

Research on the effects of dislocations on the electronic properties of semiconductors dates back to 1953. First, Shockley reported that dangling bonds at the core of dislocations in germanium (Ge) and silicon (Si) should give rise to levels lying in the forbidden band gap\cite{shockley1953dislocations}. Later, Read created a model for defect states; which assumed dislocations in Ge as acceptor type\cite{read1954lxxxvii}. Another model proposed by Schr\"{o}ter and Labusch claimed that 1D defect states could be acceptor or donor type\cite{labusch1980electrical}. Even though dislocations in simple tetrahedrally bonded semiconductors (Si, Ge etc.) have been investigated by a variety of theoretical and experimental methods\cite{broudy_electrical_1963,bell_effect_1966,patel_change_1967,alexander_dislocations_1969}, no unique and explicit model for the electronic states associated with dislocations in the band gap is adapted. For more than 40 years, most efforts to get insight into position of dislocation states in the gap resulted in ambiguity\cite{holt2007extended}. The main challenges in the analysis and interpretation of experimental data were attributed to the presence of networks of various type of dislocations, kinks and jogs, deformation induced point defects, and the ubiquitous interaction between dislocations and point defects \cite{holt2007extended,claeys2009extended}. On the other hand, theoretical calculations could not clarify inconsistent experimental data because of once-limited computational power or finite size effects. Most theoretical researchers concluded their work by highlighting that correct positioning of defect levels can only be determined using realistic Hamiltonians and large numbers of atoms in calculations \cite{jones1979theoretical,marklund1992energy,yong1989electron,lodge198990}.  

Despite the fair amount of work produced in the past, there is still a lack of complete understanding of the electronic properties of dislocations in elemental semiconductors. 

Recently, thanks to increases in computing power \cite{laukkonen2019preparing} and modern electronic density functional thoery (DFT) codes, first principles calculations of dislocations, using realistic functionals and large simulation systems are now possible\cite{pizzagalli2018first,belabbas2015electronic}. Here, we present a systematic first principles study of partial dislocations in diamond with accurate positioning of their energy levels with respect to the host crystal's band structure. The electronic band structure of dislocations as well as the anisotropic carrier mobility in directions parallel and perpendicular to the dislocation line are calculated. The results show that metallic and semiconducting dislocations arise in diamond. 1D metallic bands are revealed within the core of unreconstructed (30$^{\circ}$) partial dislocations, with a characteristic 1D density of states ($1/\sqrt{E})$. This 1D Fermi gas is spatially localized at a single-atom-wide $p_z$ orbital chain along the dislocation line. In contrast, unreconstructed pure edge dislocations in diamond are 1D semiconductors with a direct band gap of 3.21 eV. Interband transitions within the latter theoretically explain the origin of the blue band luminescence in diamond, which the literature widely report to be correlated with dislocations\cite{kiflawi_linearly_1974,pennycook_observation_1980,yamamoto_cathodoluminescence_1984,ruan_band_1992,iakoubovskii_luminescence_2000}. These results prove that it is the core states of the dislocations themselves in diamond that give rise to functional (electrical and optical) properties, rather than their distortion of the surrounding bulk diamond states. This opens the door to consider dislocations as 1D quantum phases.%somethingl like this, similar stuff in the MURI proposal?

In the rest of this paper, we first describe the computational set up. Next, we report the energetics of all calculated core configurations, as well as the ground state electronic properties of each configuration obtained using realistic hybrid functionals. Finally, carrier mobilities are calculated and discussed.

\section{COMPUTATIONAL METHODS}
\label{sec:method}

Density Functional Theory (DFT) calculations are performed with the Vienna ab initio Simulation Package (VASP)\cite{kresse1996efficient} using  projector augmented wave (PAW) pseudopotentials \cite{kresse1999ultrasoft}. Exchange correlations are treated by hybrid functionals with a Hartree-Fock mixing parameter $\alpha$ of 0.18 parametrized by Heyd, Scuseria and Ernzerhof \cite{heyd2004efficient} to eliminate the band gap underestimation problem of DFT. Incorporating 0.18 fraction of Hartree-Fock exchange recovered underestimation from 4.10 eV to 4.95 eV within the limits of computational resources. A plane wave cut-off energy of 550 eV is used with a k-points density of 0.1 \r{A}$^{-1}$ for structural relaxations. Full periodic boundary conditions are used with a quadrupolar arrangement of dislocations using a triclinic simulation cell containing two dislocations with opposite Burgers vectors.  The simulation cell consists of 576 atoms  with a separation of $\approx$ 20 \r{A} ~between the two dislocations and is oriented along \hkl[-1-12], \hkl[111] and \hkl[1-10] corresponding to the x, y, and z directions, respectively. Two sets of supercells with single and double lattice translation periods are used along the \hkl[1-10]  dislocation line direction to study core reconstruction.

The predominant slip system in the diamond cubic crystals is \hkl{111}\hkl<110>.
A dislocation is titled as glide (shuffle) when slip takes place between widely (closely) spaced $\{111\}$ planes. We only consider the glide set of dislocations since they are glissile (capable of gliding) and the most stable\cite{pizzagalli2008dislocation}. The dislocations are introduced by imposing their elastic displacement field using anisotropic elasticity theory. All atomic positions are subsequently optimized until forces are smaller than 10 meV/\r{A}$^{-1}$.

To be able to directly compare the  defective and perfect (bulk) simulation cells, the electrostatic potential in these cells need to be aligned~\cite{lany2009accurate} via 

 $$E_{VBM}= E_{VBM}^{Perfect} + V_{ave}^{Bulk-like} - V_{ave}^{Perfect}. $$

Here, $E_\text{VBM}$ and $E_\text{VBM}^\text{Perfect}$ correspond to the valence band maximum of the defective and the perfect cell respectively, $V_\text{ave}^\text{Bulk-like}$ is the average potential in the bulk like region of the defective cell and $V_\text{ave}^\text{Perfect}$ is the average potential of the perfect cell. In the defective cell, atoms with  volumetric strain values smaller than $10^{-4}$ are considered bulk-like and are used to compute $V_{ave}^{Bulk-like}$. For obtaining the band structure along the \hkl[-1-12] direction, band unfolding is performed with the fold2bloch code using 15 equidistant k points \cite{rubel2014unfolding}. 
Effective mass calculations were performed with curve fitting and the finite differences method. 

\begin{figure}

\includegraphics[width=\textwidth]{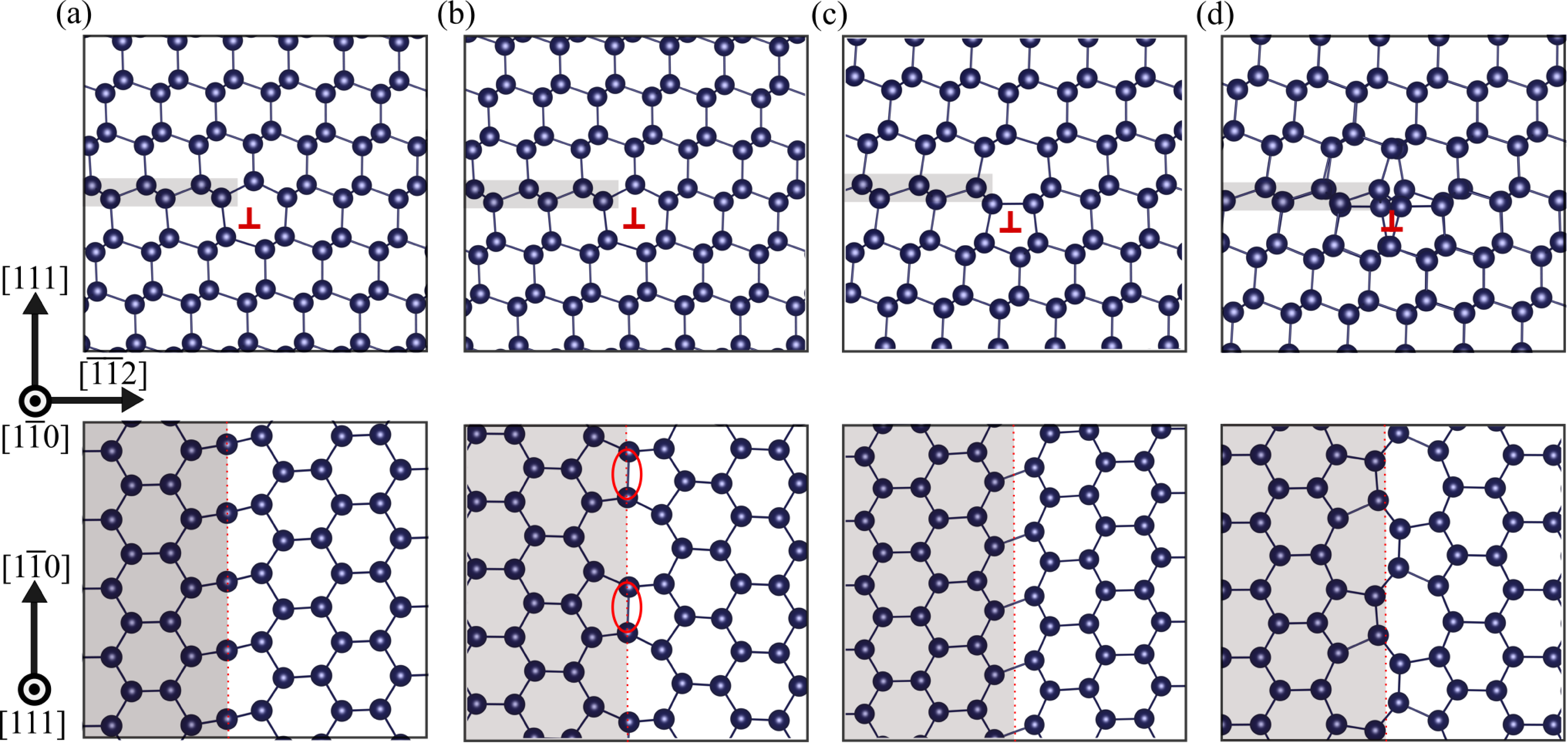}
\caption{ Relaxed dislocation core structures of the glide set of Shockley partials in diamond. The relaxed core structure of (a) Single period 30$^{\circ}$, (b) double period 30$^{\circ}$, (c) single period 90$^{\circ}$, and (d) double period 90$^{\circ}$ Shockley partial dislocations. For each structure, top figures show the top view of \hkl(1-10) plane and bottom figures show \hkl(111) glide plane. Stacking fault region associated with each partial is shaded. Location of dislocations on top figures are marked with a red colored $\perp$ sign. The new bonds formed during core reconstruction are marked by red ellipses.} Reconstructed bonds are showed in red circles for double period 30$^{\circ}$ type.
\label{fig:wide}
\end{figure}

\section{RESULTS AND DISCUSSION}
\begin{table}[b]
\caption{\label{tab:table1}
Energy difference between the reconstructed  and unreconstructed partial dislocations obtained by allowing double lattice periodicity (DP) and single lattice periodicity (SP) along the line direction.}
\begin{ruledtabular}
\begin{tabular}{lc}
 &$\left(E^\text{DP}-E^\text{SP}\right)/atom$ (\SI{}{\eV})\\
\hline
%60$^{\circ}$ Undissociated   & -0.043  \\
30$^{\circ}$ Shockley Partial& -0.046 \\
90$^{\circ}$ Shockley Partial& -0.025  \\
\end{tabular}
\end{ruledtabular}

\end{table}
 
 Previous studies~\cite{blumenau2003dislocations,blumenau2002dislocations} compared the relative stability of the glide set of dislocations in diamond and suggested that dissociation of the 60$^{\circ}$ glide dislocation into 90$^{\circ}$ and 30$^{\circ}$ Shockley partials lying on the same \hkl{111} glide plane is energetically favored. Therefore, we only investigate the electronic structure of Shockley partials with  two types of reconstruction- i.e.‘Single Period’ (SP) and ‘Double Period’ (DP)- in  this paper.
 
 \tab{table1} presents the energy per atom for different relaxed dislocation core configurations. 
The corresponding relaxed  core configurations  are shown in Fig. 1. SP-30$^{\circ}$ partial  dislocations  have dangling bonds at their core as shown in Fig. 1(a). Doubling the lattice periodicity along the dislocation line allows for  pairing dangling bonds of every second atom with their neighbors (Fig. 1(b)) and reduces the energy by 46 meV/atom. 
In case of the 90$^{\circ}$ partial dislocation, all the core atoms in both DP and SP dislocations are four-fold coordinated with highly distorted bonds as shown in Fig. 1((c)-(d)). Although no dangling bonds occur in either types of dislocations, the DP configuration is more favorable than the SP by ${\sim}$ 25 meV/atom. The bonds are up to 15\% stretched in the SP configuration and up to 11\% stretched in the DP configuration with respect to the relaxed C-C bond length (1.54 {\AA}) in perfect bulk diamond. Thus, strain in the SP core is released through the bond rearrangement, enhancing the stability of DP-90$^{\circ}$ partials. 
The small energy differences between the SP and DP glide set of dislocations imply that the structure adopted (DP or SP) can be altered depending on the environment, for example, by local strains, doping, or thermal excitation. Therefore, we consider all of these core configurations for electronic structure calculations. Next, we study the dislocation band structures and analyze the correlation between the their electronic properties and core geometries. %was investigated focusing on the analysis of induced electronic defect states and their atomic origins. 
The electronic structure of defect-free diamond in an oblique cell configuration, comparable to dislocation supercells, is used as the reference.

\begin{figure}
\includegraphics[width=8.6cm,height=8.6cm]{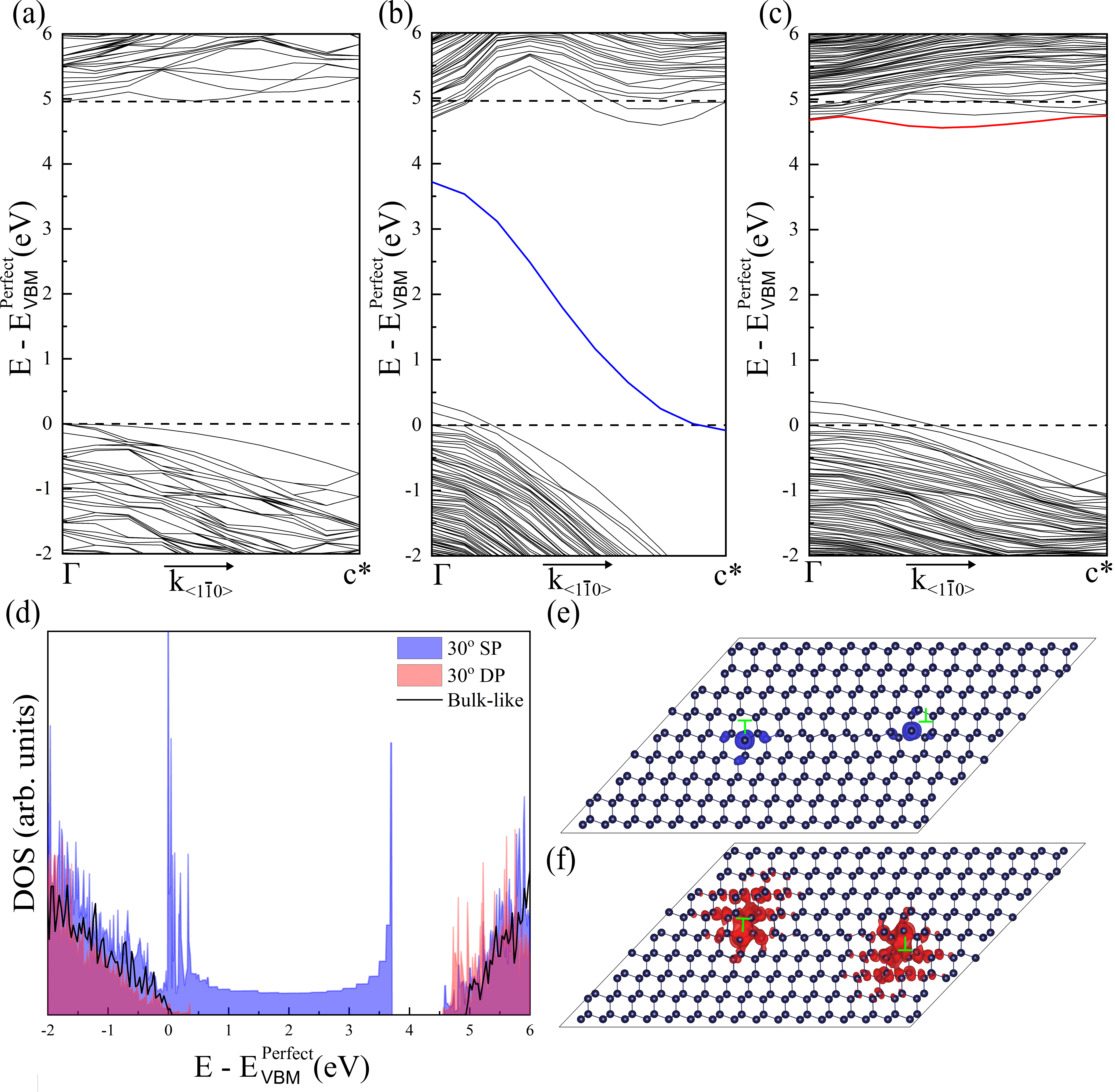}% Here is how to import EPS art
\caption{\label{fig:epsart1} Calculated electronic band structures along the dislocation line direction \hkl[1-10]  for (a) reference perfect (defect free) supercell, supercell with (b) SP 30$^{\circ}$ and (c) DP 30$^{\circ}$ Shockley partials dislocation dipoles (d)Calculated total electronic density of states for SP and DP 30$^{\circ}$ dislocation dipole supercells in reference to perfect supercell. Band decomposed charge density distributions for (e) blue colored deep defect level in band structure of SP 30$^{\circ}$ dislocation dipole supercell and (f) red colored shallow defect level in DP 30$^{\circ}$ dislocation dipole supercell. Location of dislocation dipoles on figures are marked with green colored ${\perp}$ and ${\top}$ signs. Dashed lines on each panel show the conduction band minimum and the valence band maximum of the perfect bulk diamond  as a reference.}
\end{figure}

%The electronic structure of defective supercells is aligned with that of bulk by electrostatic potential alignment method described in methods section.

Figures 2((a)-(c)) present the electronic band structures obtained along the dislocation line direction \hkl[1-10] for the reference and 30$^{\circ}$ dislocation dipole supercells. Our results reveal defect related states in DP-30$^{\circ}$ are localized and located rather close to conduction band (CB) edge in comparison to defect states in SP-30$^{\circ}$. On the other hand, the SP-30$^{\circ}$ partial gives rise to an extremely broad defect state overlapping with the valence band and extended through the gap up to the proximity of the conduction band due to a row of dangling bonds propagating along the dislocation line. These mid-gap states are found to be half filled, which implies both acceptor and donor activity is possible depending on the position of the Fermi level. Disappearance of the mid-gap states in the DP reconstruction is attributed to the elimination of these dangling bonds.

It should also be noted that two degenerate defect states are observed in each case due to the existence of two dislocations in the supercell.

Density of states (DOS) plots are shown in Fig. 2(d), revealing that the presence of dislocations gives rise to not only multiple states in the gap region but also shifts in the valence and conduction band (CB) edges for the SP-30$^{\circ}$ partial. Moreover, the atomic origins of induced defect states are investigated through band decomposed charge density analysis (Fig. 2(e) and (f)), with the defect states and corresponding partial charge densities color-coded. It is evident that these state are located in the dislocation core region, and that the degree of spacial localization for the SP-30$^{\circ}$ reconstruction is larger than that of the DP-30$^{\circ}$.

Figures 3(a)-(d) exhibit that neither the SP nor DP reconstructions in 90$^{\circ}$ partial dislocations induce mid gap states well-separated from the VB and CB edges of bulk diamond. However, the SP-90$^{\circ}$ core, with higher strain, gives rise to a relatively dispersive conduction band compared to the DP-90$^{\circ}$. Similar to the case of the 30$^{\circ}$ dislocations, the decomposed partial charge densities show that the dislocation states are localized around their core regions in real space (Fig. 3((e)-(f)). This observation is consistent with the fact that localized defect states are created by dangling bonds which are absent in either structure of the 90$^{\circ}$ partial dislocations.

\begin{figure}[b]
\includegraphics[width=8.6cm,height=8.2cm]{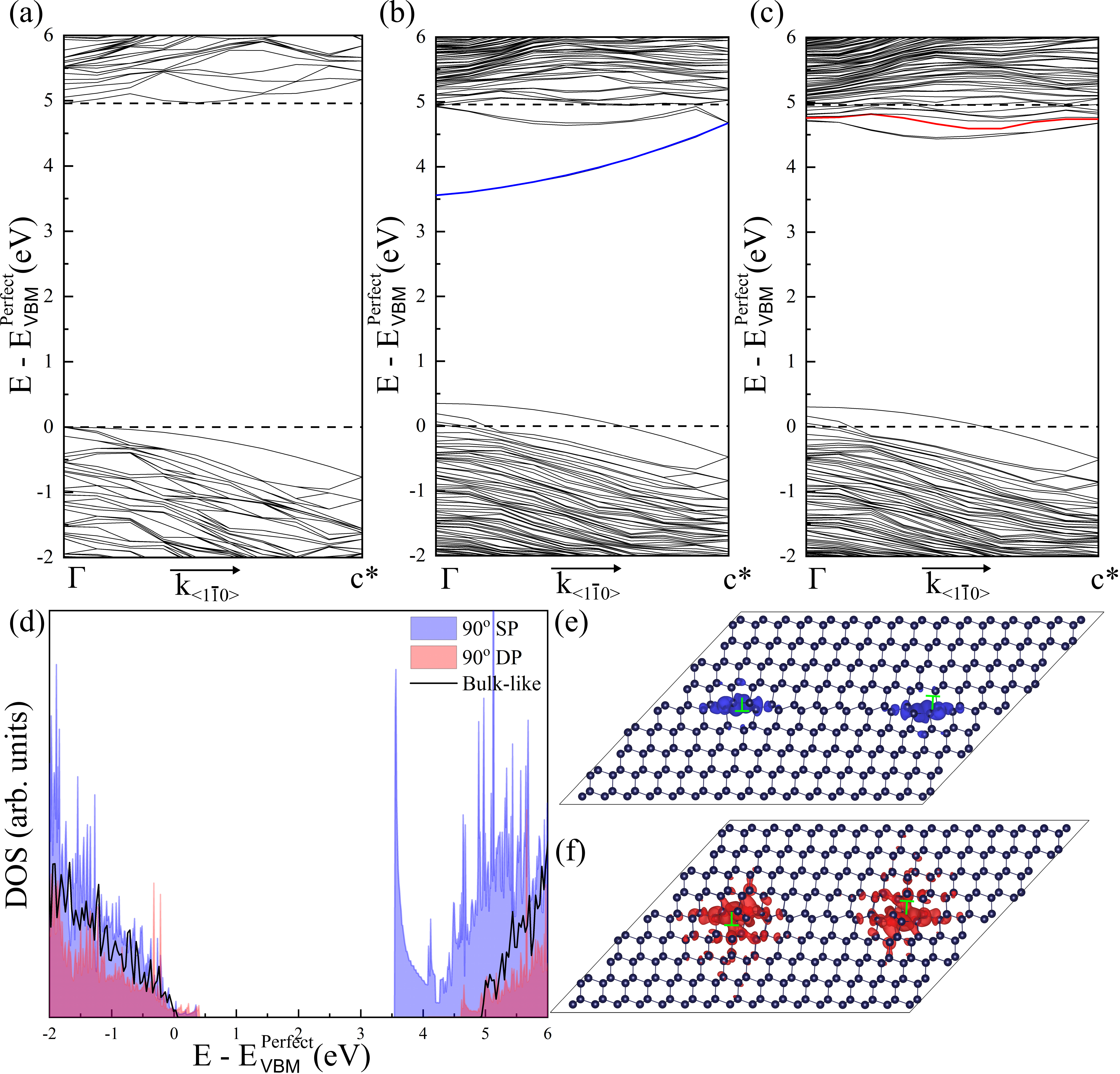}% Here is how to import EPS art
\caption{\label{fig:epsart2} Calculated electronic band structures along the dislocation line direction \hkl[1-10]  for (a) reference perfect (defect free) supercell, supercell with (b) SP 90$^{\circ}$ and (c) DP 90$^{\circ}$ Shockley partials dislocation dipoles (d) Calculated total electronic density of states for SP and DP 90$^{\circ}$ dislocation dipole supercells in reference to perfect supercell. Band decomposed charge density distributions for (e) blue colored deep defect level in SP 90$^{\circ}$ dislocation dipole supercell and (f) red colored shallow defect level in DP 90$^{\circ}$ dislocation dipole supercell. Location of dislocation dipoles on figures are marked with green colored ${\perp}$ and ${\top}$ signs. Dashed lines on each panel show the conduction band minimum and the valence band maximum of the perfect bulk diamond  as a reference.}
\end{figure}

\begin{figure}[t]
\includegraphics[width=\textwidth]{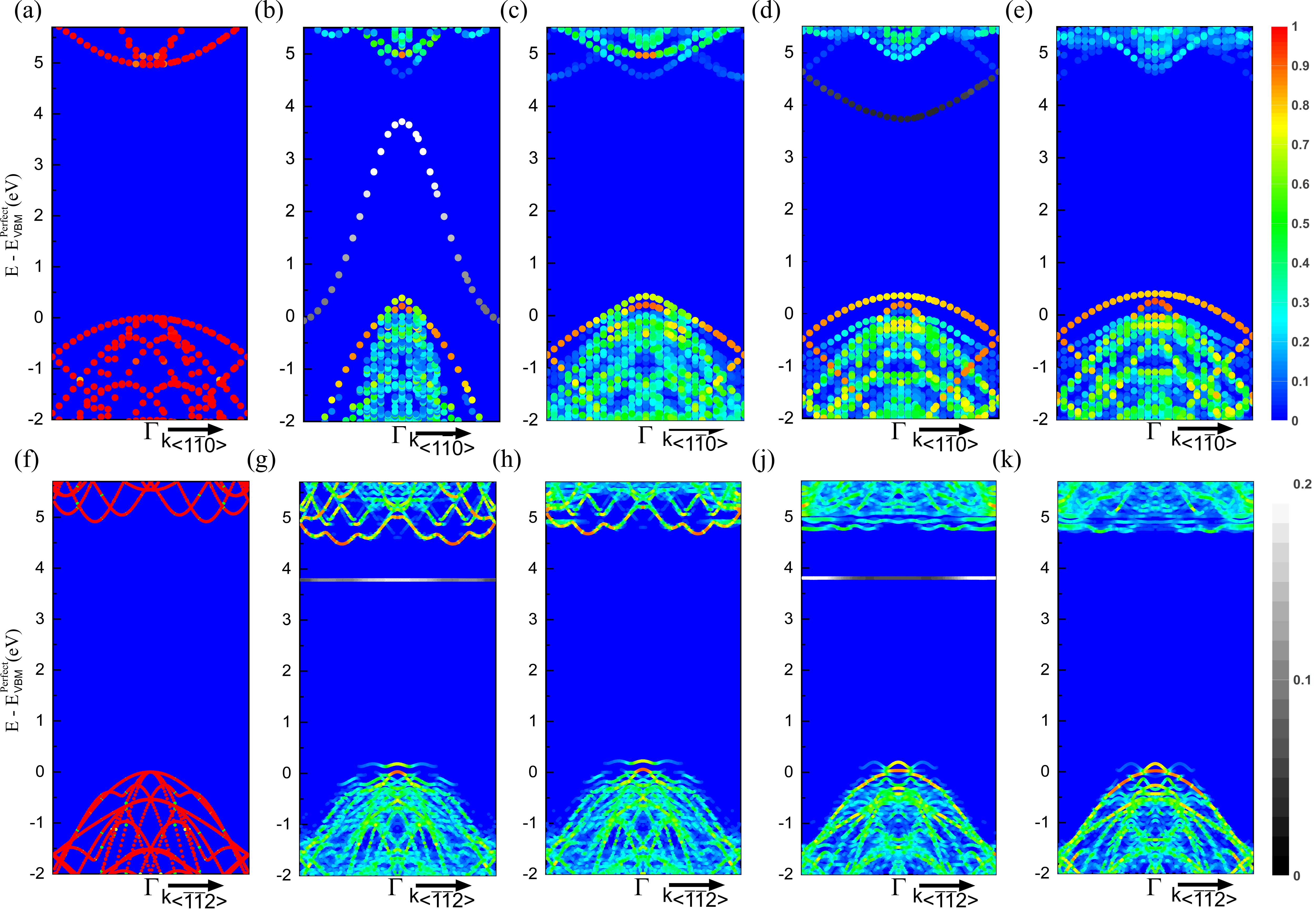}

\caption{\label{fig:wide2}Calculated electronic band structures for (a) perfect, (b) SP 30$^{\circ}$, (c) DP 30$^{\circ}$, (d) SP 90$^{\circ}$, (e) DP 90$^{\circ}$ dislocation dipole supercells, unfolded into Brillouin zone of the primitive cell using fold2bloch code\cite{rubel2014unfolding}. Top figures are the unfolded band structures along the dislocation line direction[1-10]. Bottom figures are the unfolded band structures along the Burgers vector direction \hkl[-1-12]. Color bars represent the Bloch spectral weight. Since the Bloch spectral weights of defects are in the range of [0,2], gray scale color bar (bottom) is used for defect states that are well separated from VB and CB of SP 30$^{\circ}$ and 90$^{\circ}$ dislocation dipole supercells.}
\end{figure}

Notice that the top of the valance band edges are shifted to higher energy levels for both the 30$^{\circ}$ and 90$^{\circ}$ partials in comparison to the bulk diamond reference (Fig. 2 and 3(a)-(c)). We found that these shifted bands are occupied and localized around the stacking fault region between the two dislocations. 

We also calculate the band structure along the x direction \hkl[-1-12]), which is perpendicular to the dislocation line.
Band folding effects arise in this case due to having more than one lattice translation  along the x direction of the $8\times 1 \times 1$ supercell. Note that, in the case of DP supercells, the double-periodicity is considered as the new translation vector along the line due to reconstruction.  

\fig{wide2} shows the unfolded band structures. Because, the position of the valence and conduction band edges are affected by folding along the x direction, we unfolded all the band structures for comparison.
For SP-30$^{\circ}$ partials, dispersionless electronic states (i.e. flat bands) are observed along the x direction implying confinement of carriers in real space along this direction (Fig. 4(g)). On the contrary, the half filled defect state having metallic conductivity along the line direction is quite dispersive (Fig. 4(b)).

 \begin{figure}[t]
\includegraphics[width=8.6cm,height=3.8cm]{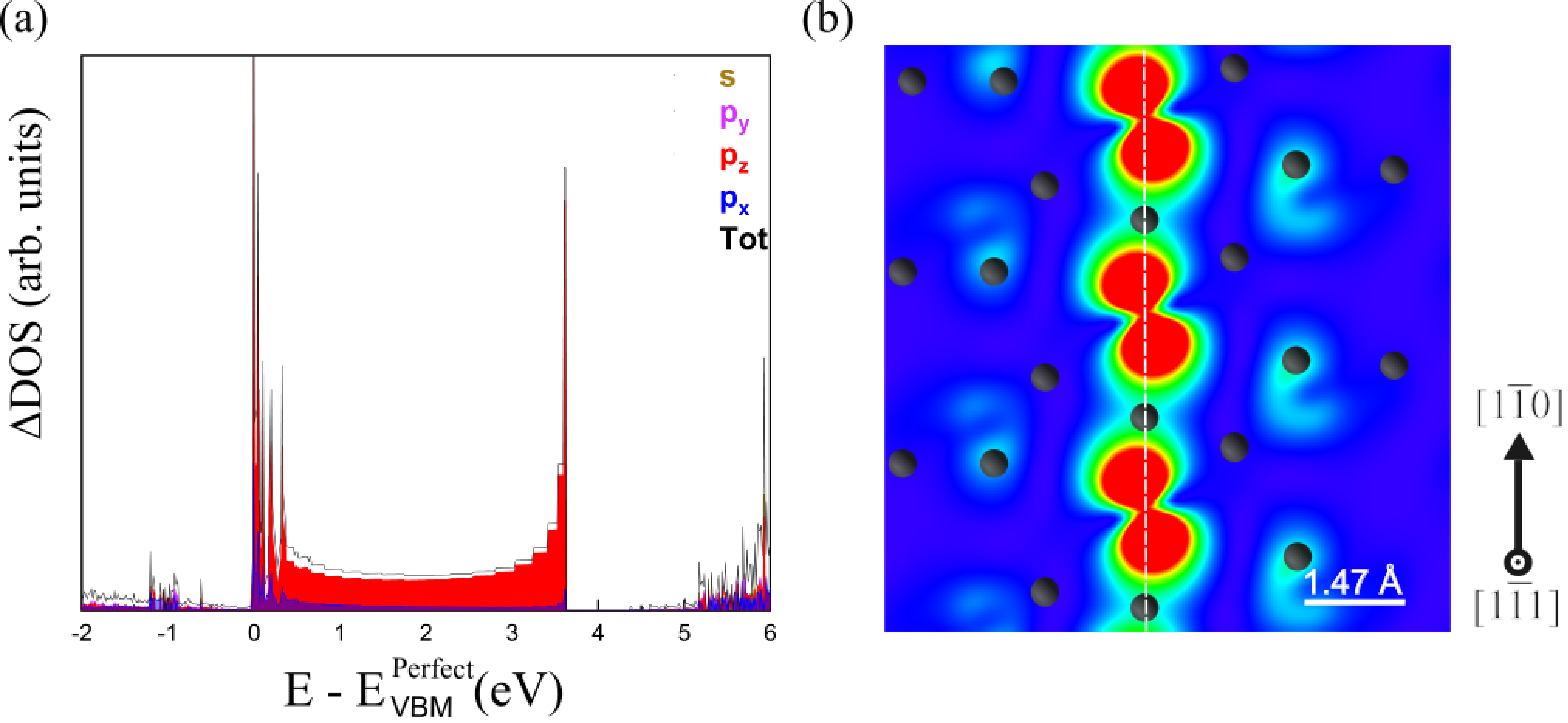}% Here is how to import EPS art
\caption{\label{fig:pz}(a) Orbital resolved partial density of states for SP 30$^{\circ}$. Inset shows the dislocation core and light blue colored atoms that are used to plot partial density of states (PDOS).  (b) Charge density plot along the dislocation core of SP 30$^{\circ}$.( Black balls represents the atoms lying on the glide plane) }
\end{figure}

\fig{pz} shows the partial DOS and the charge density plot along the dislocation line of the SP-30$^{\circ}$ partial dislocation.
It is evident that these states largely consist of $p_z$ orbitals; the charge density is localized along the line direction showing an  array of overlapping $p_z$ orbitals. Heavily localized charge density along the dislocation line combined with the dispersive band structure, demonstrate that 1D conduction takes place along the line direction within atomically narrow channels. These 1D metallic bands exhibit a DOS reduction of $1/\sqrt{E}$ at the conduction band-like bottom, and vice versa, at the valence band-like top of the band. Similar ideal 1D Fermi gases have been prevously generated using scanning tunneling microscopy (STM) to arrange metallic atoms into chains, however they degrade quickly on a surface\cite{nilius_development_2002}. Our results suggest that ideal quantum wires made from naturally occurring line defects in wide band gap materials may overcome these limitations.%something like this

Next, we quantify the carrier mobility  in the directions along and perpendicular to the dislocation line by calculating the effective masses in both directions. When there exists a localization, the effective mass approximation is not valid. However,  we have already shown that the dangling bond wavefunctions extending along the dislocation line are heavily delocalized in that direction for SP-30$^{\circ}$. Therefore, we compute the effective masses of carriers within the defect states along the line direction for SP-30$^{\circ}$ $m_{\textbf{e},defect}^{\hkl[1-10]}$ to be equal to $0.15 m_{o}$ at the $\Gamma$ point and $m_{\textbf{h},defect}^{\hkl[1-10]}$ to be equal to $0.17 m_{o}$ at the Brillouin zone edges, using a simple curve fitting as explained in \sect{method}. 

\begin{table}[h]
\label{tab:effectivem}
\caption{\label{tab:table}Calculated effective mass values in units of $m_0$ along different crystallographic directions of bulk diamond in literature and in this work}
\begin{ruledtabular}
\begin{tabular}{ccc}
 {Parameter}&{Literature (Bulk)}&{This work}\\
 \hline
$m_{hh}^{\hkl[111]}$& $0.56$~\cite{lofaas2011effective}& $0.56$\\
$m_{hh}^{\hkl[110]}$& $2.12$~\cite{amma2010electrical},$1.34,0.653
$~\cite{akimoto2014high} & $1.57$\\
$m_{hh}^{\hkl[100]}$& $0.40$~\cite{lofaas2011effective}& $-$\\
$m_{lh}^{\hkl[111]}$& $0.53$~\cite{lofaas2011effective}& $0.28$\\
$m_{lh}^{\hkl[110]}$& $0.23$~\cite{lofaas2011effective},$0.263$~\cite{akimoto2014high}& $0.20$\\ 
$m_{e}^{t}$& $0.36$~\cite{willatzen1994linear,lofaas2011effective}& $0.45$\\ 
$m_{e}^{l}$& $ 1.40$~\cite{willatzen1994linear,lofaas2011effective}& $1.09$\\ 
\end{tabular}
\end{ruledtabular}
\end{table}

\tab{effectivem} shows the calculated carrier effective masses in bulk diamond for reference, revealing that there is a far higher mobility of holes along the dislocation line direction compared to a perfect diamond crystal.
Similarly, defect states of SP-90$^{\circ}$ dislocations are delocalized along the line direction but localized along the perpendicular direction (Figs. 4(d),(j)). Effective mass of carriers within the defect states along the line direction for SP-90$^{\circ}$ is calculated as  $m_{\textbf{e},defect}^{\hkl[1-10]}$ to be equal to $1.01 m_{o}$ at the $\Gamma$ point. No  mid gap states are observed along both line and perpendicular direction for any of dislocations with DP reconstruction (Figs. 4(c),(h),(e) and (k)).

 The unfolded band structures in Fig. 4 also provide insight to the anisotropic optical properties possible in dislocated diamond. Bulk diamond (Figs. 4(a),(f)) exhibits a large indirect band gap of 4.95 eV. 
 However, in the case of pure edge dislocations, (SP-90), a much smaller direct band gap of 3.21 eV is seen in our calculations. This implies the onset of optical absorption at 3.21 eV in dislocated diamond, and likely, luminescent emission of photons at, or slightly below this energy assuming some excitonic (e-h) coupling occurs. The literature on the optical properties of diamond is quite extensive, and it is widely reported that the broad blue band emission in diamond (centered at 2.8 eV) arises from dislocations\cite{kiflawi_linearly_1974,pennycook_observation_1980,yamamoto_cathodoluminescence_1984,ruan_band_1992,iakoubovskii_luminescence_2000}. Cathodoluminescence (CL) studies show blue emission in diamond arising exactly at the location of dislocations. Although CL can localize the luminescence to the vicinity of dislocations, it was not clear if radiative recombination was occuring in the core dislocation states themselves, or in the bulk diamond states around the dislocations. The latter could occur, for example, if point defects, or strain-generated carrier trapping and recombination pathways develop near the dislocations. CL excites a large population of hot carriers, well above the band edges and as a result, cannot determine the absorption onset of the blue emission. But later, Iakoubovskii and Adriaenssens examined photoluminescence excitation (PLE) spectroscopy and identified an absorption onset for the blue band emission centered at 3.0 eV. These results demonstrated that excitation of photocarriers in the bulk-states of diamond ($\approx${5.5} eV) is not needed to drive the blue band emission. The band structure in Fig. 4(d) provides a straightforward explanation for the blue band CL and PLE emission data, indicating that the 2.8 eV emission band arises from band-to-band recombination in the core states of pure edge dislocations in diamond and with an absorption onset at 3.21 eV. We note that, our calculated band gap values underestimate the experimental value, eg 4.95 eV compared to $\approx${5.5} eV for bulk diamnod. Therefore, while the numbers cannot be directly compared to the experimental measurements, the trend is in good agreement with the optical literature on diamond. Additionally, the anisotropy of the band structure also suggests an optical polarization axis with blue band emission parallel to the dislocation line, i.e. electric-dipole transitions along the dispersing band k-vector. Again, literature backs up this prediction with CL data of Kiflawi and Lang demonstrating $>90\%$ polarization of blue band emission parallel to dislocation lines.

\section{CONCLUSIONS}

Ground state electronic properties of the glide set of partial dislocations in diamond were calculated. 
We found that (1) the position and effective mass of dislocation-induced states depend heavily on the core structure, (2) only mixed dislocations with dangling bonds have metallic conductivity through half-filled gap states, (3) 1D conduction along the line direction of these dislocations is attributed to a chain of overlapping $p_z$ orbitals forming a dispersive band along the line defect, (4) Pure edge dislocations in diamond exhibit a direct band gap of 3.21 eV providing a theoretical explanation on the origin of the blue band emission in diamond, and (5) the DOS for the core states is that of an ideal 1D Fermi gas in both metallic and semiconducting dislocations. Consequently, dislocations with under coordinated core atoms appear to be naturally formed 1D quantum wires. Core states of dislocations in wide band gap materials like diamond  could be used as an active components in functional materials.

In this study, the geometry and electronic properties of ideal (straight and clean) partial dislocations are examined. Jogs, kinks, dislocation nodes, point defect decorated dislocations can  alter the effects on electronic structure; which should be investigated thoroughly in future.

\begin{acknowledgments}
We gratefully acknowledge the support of this work by the AFOSR grant number FA9550-21-1-0278. Computational resources were provided by the Ohio Supercomputer center.
\end{acknowledgments}

\bibliographystyle{apsrev4-2}
\bibliography{dislocation}

\pagebreak

\end{document}